\documentclass[aps,prb,notitlepage,showkeys,twocolumn,superscriptaddress]{revtex4-2}

\usepackage{amsmath}
\usepackage{amssymb}
\usepackage{graphicx}
\usepackage{epstopdf}
\usepackage[colorlinks=true]{hyperref}
\usepackage{float}
\usepackage[normalem]{ulem}

\begin{document}

\preprint{APS/123-QED}

\title{
Effects of homogeneous doping on the electron-phonon coupling in SrTiO$_3$}

\author{Minwoo Park}
\affiliation{Department of Physics and Natural Science Research Institute, University of Seoul, Seoul 02504, Republic of Korea}
\author{Suk Bum Chung}
\affiliation{Department of Physics and Natural Science Research Institute, University of Seoul, Seoul 02504, Republic of Korea}
\affiliation{School of Physics, Korea Institute for Advanced Study, Seoul 02455, Korea}

\date{\today}

\begin{abstract}
Bulk n-type SrTiO$_3$ (STO) has long been known to possess a superconducting ground state at an exceptionally dilute carrier density. This has raised questions about the applicability of the BCS-Eliashberg paradigm with its underlying adiabatic assumption. However, 
recent experimental reports have set the  pairing gap to the critical temperature ($T_c$) ratio at the BCS value for the superconductivity in Nb-doped STO, even though the adiabaticity condition the BCS pairing requires is satisfied over the entire superconducting dome only by the lowest branch of optical phonons. In spite of the strong implications these reports have on specifying the pairing glue, they have not proved sufficient in explaining the magnitude of the optimal doping. This has motivated us to apply the density functional theory to Nb-doped STO to analyze how the phonon band structures and the electron-phonon coupling evolves with doping. To describe the very low doping concentration, we 
tune the homogeneous background charge, 
from which we have obtained a 
first-principles 
result on the doping dependent phonon frequency that is in good agreement with experimental data for Nb-doped STO. Using the EPW code, we obtain the doping dependent phonon dispersion and the electron-phonon coupling strength. Within the framework of our calculation, we find that the electron-phonon coupling forms a dome in a doping range 
lower than 
the experimentally observed superconducting dome of the Nb-doped STO. Additionally, we have examined the doping dependence of both the orbital angular momentum quenching in the electron-phonon coupling and the phonon displacement correlation length 
and have found the former to have strong correlation with our electron-phonon coupling in the overdoped region.
\end{abstract}

\keywords{electron-phonon coupling, electron pairing interaction, polar phonon modes, doped quantum paraelectric, dilute superconductor}
\maketitle


\section{\label{sec:intro}Introduction}

One overarching theme of the STO physics 
is the effect of its proximate continuous ferroelectric (FE) transition \cite{KMuller1979, Haeni2004}. 
The 
effect is quite standard and well understood for some standard aspects, such as the high dielectric constant 
and 
the 
polar soft phonon mode 
\cite{Yamada1969, Bauerle1980, Vogt1981, Kamaras1995}. More recently, there has been much debate on 
the possible relation between the proximate FE transition and the superconducting mechanism \cite{Rowley2014, Edge2015, Rischau2017, Swartz2018, Tomioka2019, Gastiasoro2020, Yoon2021, Setty2022, Rischau2022, Yu2022, Volkov2022, Tomioka2022, Klein2023} of 
the dilute n-type bulk \cite{Schooley1964, Koonce1967}. This superconductivity has long retained its notoriety for both being the first discovered instance to exhibit the doping-dependent superconducting dome and occurring at a carrier density lower than any other bulk superconductor save the recently discovered doped Bi \cite{Prakash2017}. Given such prominent unconventional features, the results of the optical conductivity measurement \cite{Thiemann2018} and the tunneling spectroscopy measurement \cite{Swartz2018, Yoon2021} for the Nb-doped STO superconducting phase in the last few years have been quite surprising, 
revealing the pairing gap to 
$T_c$ ratio 
to be at the BCS value for most of the superconducting dome. They raise the possibility of 
STO superconductivity, with all its unconventional features, 
arising from a BCS-type pairing mechanism, which is understood to impose the adiabatic criterion, {\it i.e.} the pairing interaction whose frequency is much smaller than the electron Fermi energy. 
A pairing interaction that can intrinsically satisfy the adiabatic criterion is 
provided by the proximity to a continuous FE transition, where 
the lowest 
polar phonon mode, called the TO1 mode, softens. In the case of STO, it has been known that 
this is the only optical phonon mode 
that satisfies the adiabatic criterion throughout the superconducting dome \cite{Yoon2021}.

While the phonon mediated BCS pairing scenario 
offers some key qualitative explanations for the 
STO superconducting dome, 
it is not intrinsically sufficient in 
explaining 
the observed value of the optimal doping. 
It is true that within this scenario the pairing suppression in the overdoped region can be explained  \cite{Yoon2021,Yu2022, Gastiasoro2022} from the experimentally observed hardening of the TO1 phonons with the increasing doping concentration \cite{Bauerle1980,vanMechelen2008}, 
but understanding how the optimal doping occurs at a relatively dilute doping and thus limits the optimal critical temperature remains a challenge. 
Hence, the simplest models with the electronic coupling to the TO1 phonons 
\cite{Yu2022, Gastiasoro2023} 
sought to address the emergence of the $T_c$ dome structure rather than to make good order-of-magnitude estimates for the optimal doping. 

\begin{figure}[t]
\centering
    \includegraphics[width=.5\textwidth]{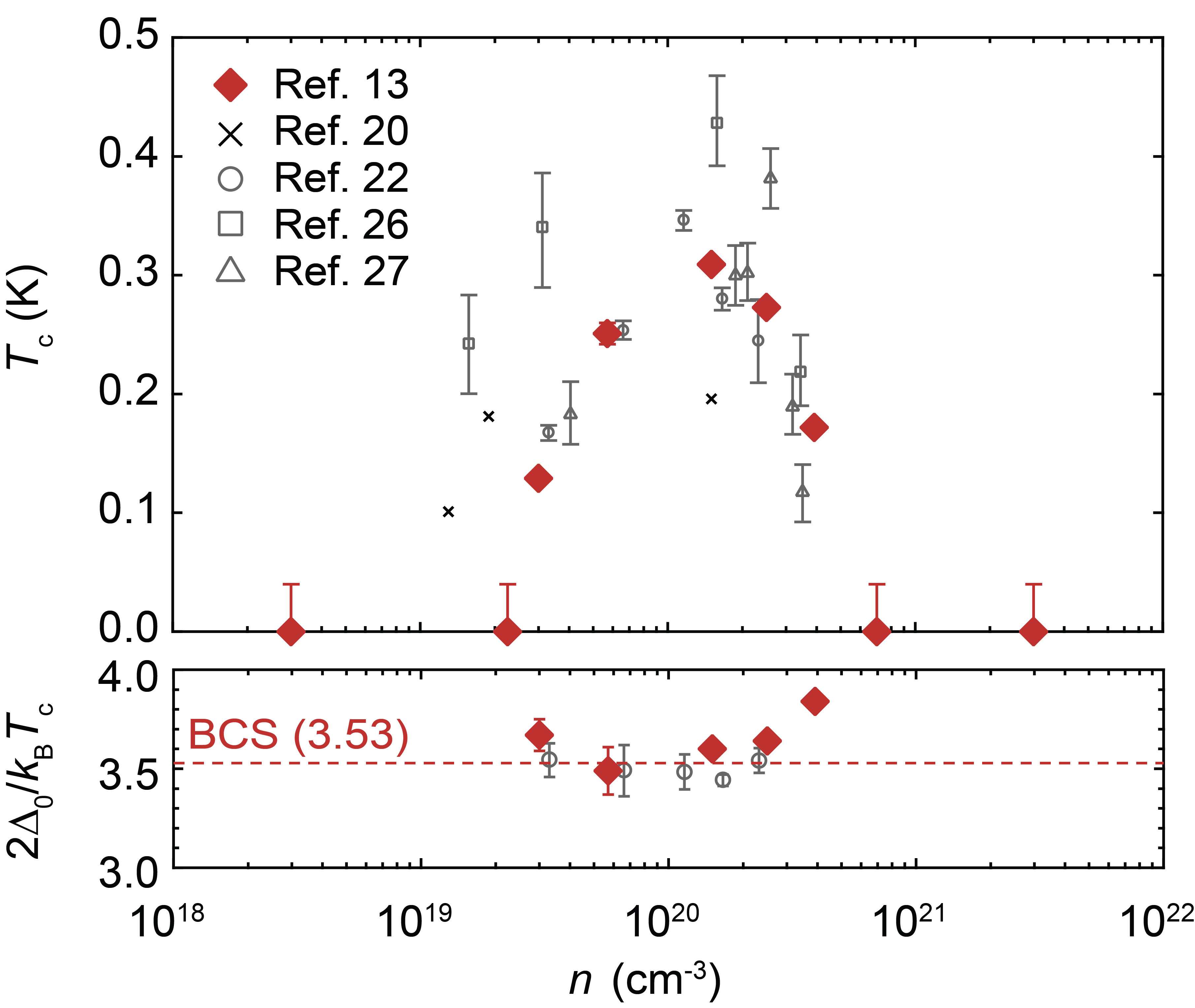} 
    \caption{Superconducting critical temperatures and pairing gaps as a function of doping concentration for Nb-doped STO from Ref.~\cite{Yoon2021}. 
    The top plot shows the resistive transitions 
    \cite{Koonce1967, XLin2014, Collignon2017, Thiemann2018, Yoon2021}, while the horizontal dashed red line of the bottom plot indicates the BCS value 2$\Delta_0$/k$_B$T$_c$ = 3.53.}
    \label{FIG:Tc}
\end{figure}

Within the framework of the phonon mediated BCS pairing scenario, 
different theories have been proposed for the mechanism that limits the optimal doping. 
According to one recent theoretical proposal \cite{Gastiasoro2023}, this suppression 
can be attributed to attenuation of the effective electron-phonon matrix element from the orbital angular momentum (OAM) quenching of the Ti $t_{2g}$ orbitals at higher doping concentration. 
In this proposal, the doping evolution of the 
phonon dispersion 
is relevant only for increasing the minimum frequency. In contrast, a recent inelastic neutron scattering experiment on n-type STO led to the suggestion \cite{Fauque}  that the superconducting critical temperature $T_c$ is closely correlated to $k_F \ell_0$, where $\ell_0$ is the 
TO1 phonon displacement correlation length and $(k_F \ell_0)^3$ therefore would be proportional to the number of electrons within the range of the TO1 phonon-mediated interaction. 
These two proposed mechanisms are not equivalent, despite both arising from the phonon-mediated pairing scenario. The former proposal is strongly dependent on the orbital symmetry of the Ti $t_{2g}$ orbital, but the phonon dispersion does not feature beyond the frequency minimum; for the latter proposal, the exact converse is the case. Finally it should be pointed out that the latter proposal 
does not have any particular dependence on the form of the electron-phonon coupling. 

Given the need for a better understanding of the mechanism for limiting the optimal doping within the framework of phonon-mediated superconductivity, first-principle calculation of both the phonon band structure and the electron-phonon coupling in STO is a natural and necessary step in 
developing the full theory of its superconductivity. 
Indeed, the 
first-principles calculations of the doping dependent phonon band structure and electron-phonon coupling have been carried out in KTaO$_3$ \cite{esswein2023firstprinciples}, another cubic perovskite transition metal oxide that is in the proximity of the continuous ferroelectric transition. We report in this paper both the phonon mode resolved linear electron-phonon coupling $\lambda_{\bf q,\nu}$ and its summation over the first Brillouin zone (BZ) $\lambda$, obtained using the 
first-principles calculation based on the density functional theory for cubic SrTiO$_3$ over the range that covers the entire experimentally reported superconducting dome. The main findings of our calculations are that i) we 
find a well-defined dome for $\lambda$ 
with an optimal doping 
about one order of magnitude smaller than 
that of the 
superconducting dome; ii) the $\lambda$ suppression in the overdoped region 
can be closely tracked 
with the OAM quenching, 
which can be expected from the 
linear 
coupling of electrons to the polar phonon mode  \cite{Gastiasoro2020, Yu2022, Gastiasoro2022, Gastiasoro2023};  
iii) $k_F \ell_0$ 
suppression is not evident on the overdoped side where $\lambda$ is strongly suppressed.

\section{Method}

Using density functional theory (DFT), we performed calculations as implemented in the Quantum ESPRESSO package v.7.1 \cite{Giannozzi2009, Giannozzi2017, Giannozzi2020}. We have employed full relativistic local density approximation (LDA) using the Perdew-Zunger (PZ) parameterization for the exchange-correlation energy functional with the projector augmented wave method (PAW) in pslibrary \cite{Perdew1981,Blochl1994,Corso2014}. The kinetic energy cutoff for wavefunctions was 60 Ry (816 eV). The 1st BZ integration was performed using the Monkhorst-Pack scheme with 16 $\times$ 16 $\times$ 16 k-point sampling \cite{MP1976}. Geometric optimization was carried out until the Hellmann Feynman force acting on each atom was smaller than 0.1 meV/\AA, from which we find the STO lattice constant to be well approximated by $a_0=$3.8565 \AA~for the three lowest doping values; for the higher doping values where the results were much less sensitive to the lattice constant we approximated the lattice constant by  $a_0=$3.8600 \AA~(see Appendix C for details). We achieved the doping effect with the background charge (so-called `jellium') method using Gaussian smearing of 0.1 meV width for five different doping levels in the range from 0.0001 to 0.02 electrons per unit cell (e/u.c.), which covers all but the upper critical doping of the experimentally measured superconducting dome \cite{Gastiasoro2020A, Yoon2021}. The phonon dispersion was calculated on a 4 $\times$ 4 $\times$ 4 q-point mesh; 200 q-points were employed between high-symmetry points. We note that our phonon dispersion calculation considers doping far more dilute than some of recent first-principles calculations, {\it e.g.} Ref.~\cite{Cancellieri2016} where the VASP code was used.



The electron-phonon coupling effect was calculated with the EPW 5.5 code in the Quantum ESPRESSO package \cite{Giustino2007, Ponce2016}. The relevant electronic bands in SrTiO$_3$ are the three Ti-$3d$ $t_{2g}$ bands, which are described by maximally localized Wannier functions using the Wannier90 code \cite{Pizzi2020} within the EPW code, including the Ti spin-orbit coupling, for which we obtain $\xi$=20meV as shown in Appendix A. The electron-phonon matrix was calculated on coarse 16 $\times$ 16 $\times$ 16 k-points and 4 $\times$ 4 $\times$ 4 q-points and then interpolated onto a fine grid. We use mixed two fine q-points meshes for efficient calculation to obtain the electron-phonon coupling constant $\lambda$ near the $\Gamma$ point. This special q-mesh is a combination of two grids, 11 $\times$ 11 $\times$ 11 grid including the $\Gamma$ point range from -0.5 to 0.5 in units of 2$\pi$/$a_0$ and 20 $\times$ 20 $\times$ 20 Monkhorst-Pack grid range from -0.1 to 0.1 in units of 2$\pi$/$a_0$. For the mode-resolved electron-phonon coupling strengths, 200 q-points were again employed between each cubic high-symmetry point. The well-known splitting between the longitudinal (LO1) and transverse (TO1) branches of the lowest optical phonon is expected to evolve, possibly attenuate, with doping due to its effect on charge screening \cite{esswein2023firstprinciples}. To the best of our knowledge, a method to compute electron-phonon coupling while obtaining the Born effective charges and dielectric constants needed for LO-TO splitting in doped insulators has not been implemented. We leave this as a future challenge and neglect the LO1/TO1 splitting altogether.






\begin{figure}[t]
    \centering
    \includegraphics[width=.5\textwidth]{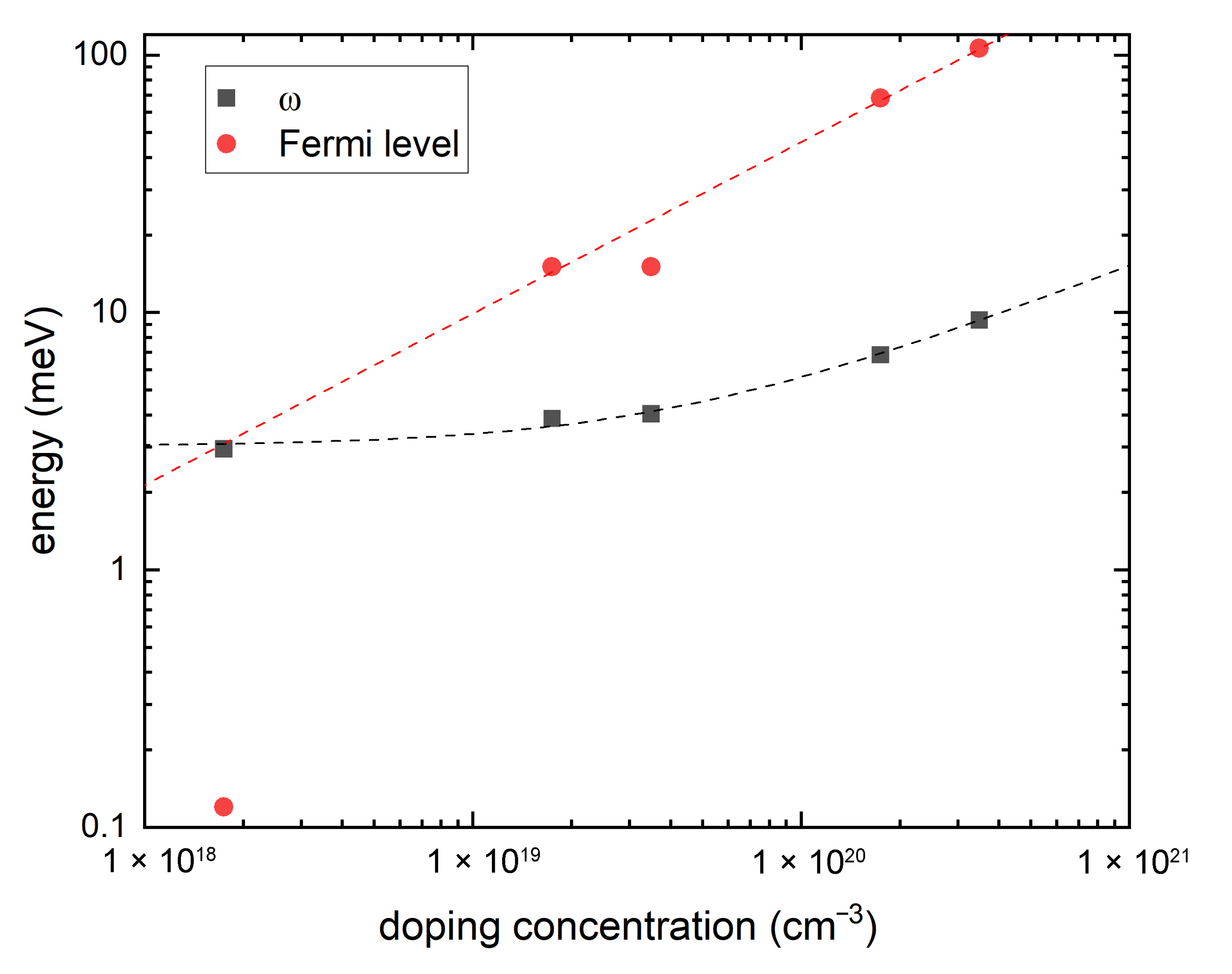} 
    \caption{Doping dependence of the polar soft mode energy at $\Gamma$ (black dots) and the Fermi energy (red dots). The black dotted curve shows the $\omega^2(n) = \omega_0^2 + \gamma_n n$ fitting and the red dotted curve the $E_F = \hbar^2(3\pi^2 n)^{2/3}/2m^*$, with $m^*$ being approximately 1.65 times the free electron mass.}
    \label{FIG:w2Evol}
\end{figure}

\section{Results}

\begin{figure*}
\includegraphics[width=.85\textwidth]{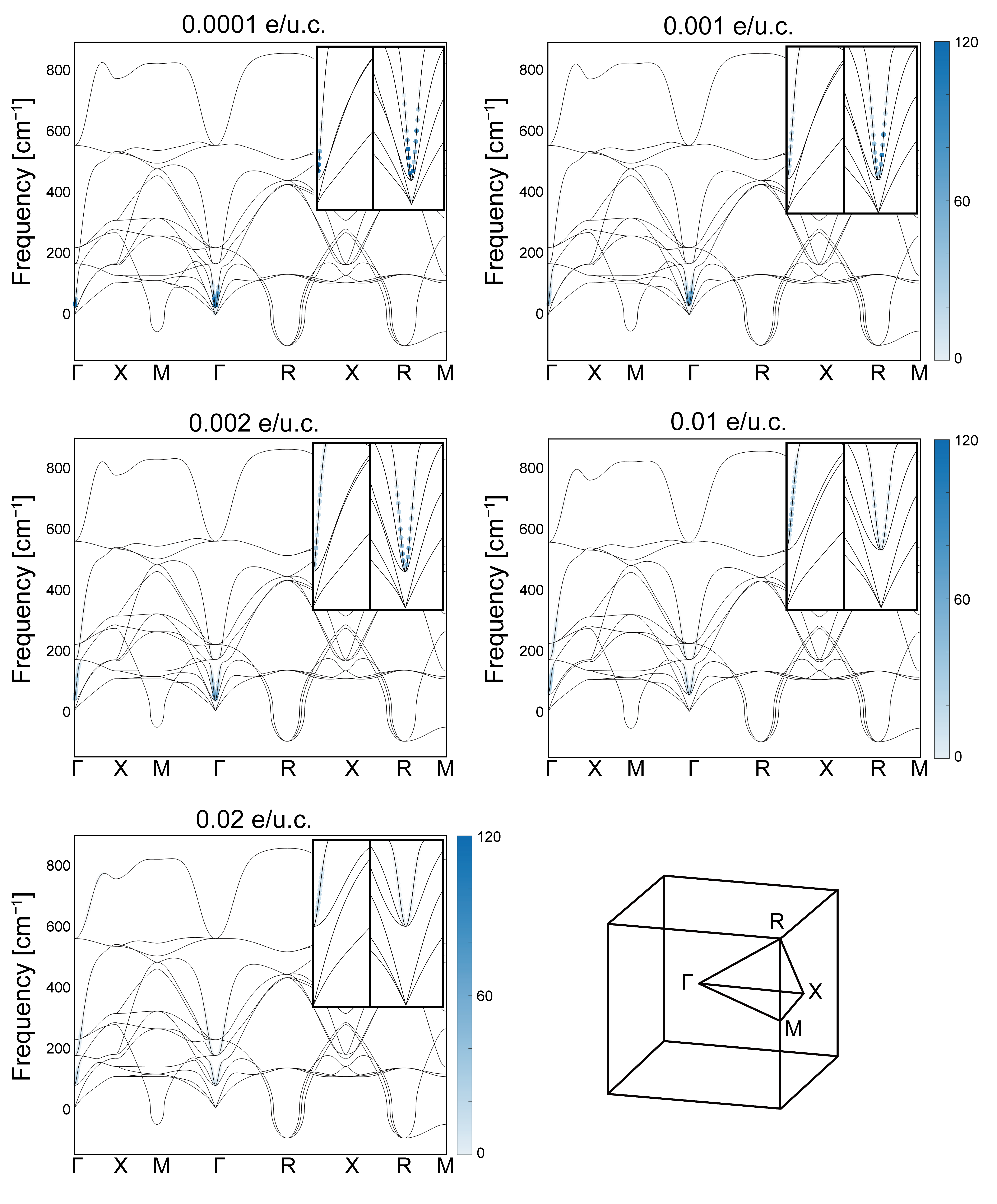}
\caption{Phonon-mode-resolved electron-phonon coupling $\lambda_{{\bf q},\nu}$ along the high symmetry lines of the 1st BZ at doping concentrations ranging from 0.0001e/u.c. to 0.02e/u.c., {\it i.e.} approximately 1.7$\times 10^{18}$cm$^{-3}$ to 3.4$\times 10^{20}$cm$^{-3}$, plotted on top of phonon frequencies (drawn in narrow black curves); note that for the frequency 1cm$^{-1}$ corresponds to approximately 0.124meV. All plots are on the same scale and share the same colorbar for the dimensionless $\lambda_{{\bf q},\nu}$. The insets in the top right corners of each subplot show a zoomed-in part of the area around $\Gamma$ towards the X, M and R points.}
\label{FIG:MRLambda}
\end{figure*}

\subsection{Phonon-mode-resolved electron-phonon coupling}

We first give a summary of the standard formalism \cite{Allen1972, Giustino2007} we have used to calculated the mode-resolved linear electron-phonon coupling strength, the key ingredient of which is the imaginary part of the phonon self-energy of the branch $\nu$ at the momentum ${\bf q}$ to the one-loop order due to electron-phonon coupling:
\begin{eqnarray}
\Pi''_{{\bf q},\nu} &=& 2{\rm Im}\sum_{m,n,{\bf k}} |g^{\rm SE}_{mn,\nu} ({\bf k},{\bf q})|^2\frac{f_{n,{\bf k}}-f_{m,{\bf k}+{\bf q}}}{\epsilon_{m,{\bf k}+{\bf q}}-\epsilon_{n,{\bf k}}-\omega_{{\bf q},\nu}-i\delta}\nonumber\\
&=& 2\pi \sum_{m,n,{\bf k}}|g^{\rm SE}_{mn,\nu} ({\bf k},{\bf q})|^2 (f_{n,{\bf k}}-f_{m,{\bf k}+{\bf q}})\nonumber\\
&\,&\times\delta(\omega_{{\bf q},\nu}+\epsilon_{n,{\bf k}}-\epsilon_{m,{\bf k}+{\bf q}}),
\label{EQ:phononISE}
\end{eqnarray}
where $\omega_{{\bf q},\nu}$ is the phonon frequency, $\epsilon_{n,{\bf k}}$ the electronic band energy for the band $n$ at the momentum ${\bf k}$. The electron-phonon matrix element here is given by 
\begin{equation}
g^{\rm SE}_{mn,\nu} ({\bf k},{\bf q})=\sqrt{\frac{\hbar}{2m_0 \omega_{{\bf q},\nu}}} \langle m,{\bf k}+{\bf q}| \partial_{{\bf q},\nu} V | n, {\bf k} \rangle.
\end{equation}
where $m_0$ is the effective mass of the phonon mode, $| n, {\bf k} \rangle$ the electronic Bloch state, and $\partial_{{\bf q},\nu} V$ the derivative of the self-consistent ionic potential $V$ with respect to the collective ionic displacement of the phonon mode ${\bf q}, \nu$. The electron-phonon coupling strength for the phonon mode ${\bf q}, \nu$ is proportional to the Eq.~\eqref{EQ:phononISE}:
\begin{equation}
\lambda_{{\bf q},\nu}=\frac{2}{\pi N_F}\frac{\Pi''_{{\bf q},\nu}}{\omega_{{\bf q},\nu}^2}
\end{equation}
where $N_F$ is the electronic density of states at the Fermi level. The electron-phonon coupling strength is simply the summation of $\lambda_{{\bf q},\nu}$ over all phonon modes and the entire 1st BZ, {\it i.e.} $\lambda = \sum_{{\bf q},\nu} \lambda_{{\bf q},\nu}$.

\begin{figure}[t]
    \centering
    \includegraphics[width=.5\textwidth]{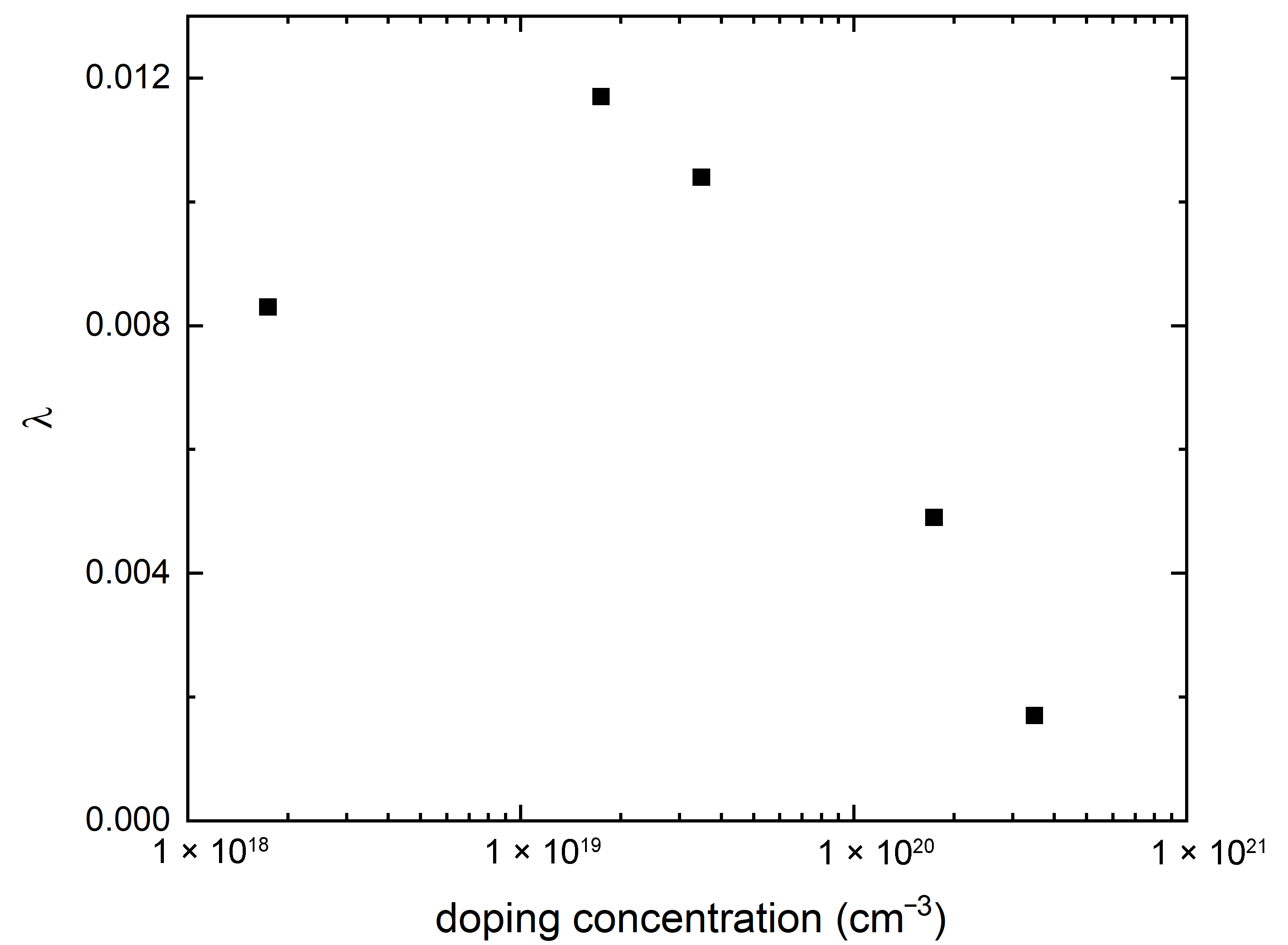} 
    \caption{The doping evolution of the total  electron-phonon coupling strength $\lambda=\sum_{{\bf q},\nu}\lambda_{{\bf q},\nu}$, the summation carried out over all phonon modes and the entire 1st BZ.} 
    \label{FIG:EPhLambda}
\end{figure}

Given that $\lambda_{{\bf q},\nu}$ can meaningfully quantify the electronic interaction mediated by phonons, {\it i.e.} without any vertex correction, only when the modes are adiabatic compared to the Fermi energy, we plot in Fig.~\ref{FIG:w2Evol} the doping evolution of the frequency of the polar soft mode at $\Gamma$ together with the Fermi energy. Both are obtained from our 
first-principles calculations, with the latter determined to be the energy at which the electron occupation number falls to 1\% of the occupation number at $\Gamma$. We see that our polar soft mode at $\Gamma$ hardens to 9.35 meV for 0.02 e/u.c.   from 2.95 meV for 0.0001 e/u.c., and this can be fitted to $\omega_{0,\nu}^2(n) = \omega_0^2 + \gamma_n n$ where $\gamma_n$= 2.2$\times 10^{-19}$meV$^2 \cdot$cm$^3$ and $\omega_0^2$ = 9.1 meV$^2$; for comparison, applying the same fitting to  experimental phonon data is is known to give $\gamma_n$=1.8$\times 10^{-19}$meV$^2 \cdot$cm$^3$ and $\omega_0^2$ = 1 meV$^2$ at 4K \cite{Bauerle1980, Gastiasoro2020A}. Meanwhile, we find that our results for the Fermi energy fit very well to the free electron formula except for the lowest doping at 0.0001 e/u.c., or equivalently, $n=1.7 \times 10^{18}$cm$^{-3}$; the discrepancy here may be attributed to the finite mesh size causing worse problems when $E_F$ becomes very small. In general, we find that the polar soft mode can be taken to be adiabatic for the doping concentration $n > 10^{19}$cm$^{-3}$. The experimentally measured superconducting dome, as shown in Fig.~\ref{FIG:Tc}, has the lower critical doping larger than $10^{19}$cm$^{-3}$, which means that the relevant underdoped region is well within the adiabatic regime. However, we have also carried out calculations for the doping of 
$n=1.7 \times 10^{18}$cm$^{-3}$; as a doping value outside the adiabatic regime, it can be used as a negative benchmark, {\it i.e.} to examine what happens to our 
first-principles calculation when the adiabaticity condition breaks down. 

We show in Fig.~\ref{FIG:MRLambda} the mode-resolved electron-phonon coupling $\lambda_{{\bf q},\nu}$, together with the full phonon spectra, along the high-symmetry direction of the 1st BZ for five different doping values. We first note that our geometrical optimization removed the instability toward uniform polar distortion, leaving no imaginary frequency near $\Gamma$ at any doping values; {\it cf.} Ref.~\cite{Cancellieri2016}. For all doping levels, we emphasize that the $\lambda_{{\bf q},\nu}$ distribution is massively concentrated in the polar soft mode near $\Gamma$. Although there is a secondary contribution from the higher-energy optical mode around $\Gamma$, we have not found any significant contribution from either the acoustic mode across the entire 1st BZ or any modes around any other high-symmetry points in the 1st BZ. We therefore argue that instabilities around high-symmetry points other than $\Gamma$ can be ignored for our purpose.

\subsection{Analysis of electron-phonon coupling}

The electron-phonon coupling strength $\lambda$, 
obtained by integrating $\lambda_{{\bf q},\nu}$ over the entire 1st BZ, is 
shown as a function of the doping concentration in Fig.~\ref{FIG:EPhLambda}. A well-defined dome structure is discernible in the calculated $\lambda$ with the optimal doping occurring around  $n=1.7 \times 10^{19}$/cm$^3$; for comparison, the optimal doping for $T_c$ in the Nb-doped STO \cite{Thiemann2018, Yoon2021} occurred around $n=1.5\times10^{20}$cm$^{-3}$, as shown in Fig.~\ref{FIG:Tc}. Closely related to the optimal doping is the value of the upper and lower critical dopings. The data in Fig.~\ref{FIG:EPhLambda} can be reasonably extrapolated to obtain $n \approx 5 \times 10^{20}$cm$^{-3}$ for the $\lambda$ upper critical doping, which can be compared to the $T_c$ upper critical doping of approximately $n = 7 \times 10^{20}$cm$^{-3}$ \cite{Yoon2021} shown in Fig.~\ref{FIG:Tc}. Evidently, there is better agreement for the upper critical doping compared to the optimal doping, even if the discrepancy in the latter is still within an order of magnitude. In contrast, the $\lambda$ lower critical doping falls outside the doping range of Fig.~\ref{FIG:EPhLambda}, which is a more marked discrepancy with the $T_c$ lower critical doping of $n = 2.5 \times 10^{19}$cm$^{-3}$. Indeed, we note here that 
due to 
the $\lambda$ optimal doping 
being considerably smaller than the $T_c$ optimal doping, 
the $\lambda$ dome does not fall entirely within the adiabatic regime, as can be seen from Fig.~\ref{FIG:w2Evol}.      

\textcolor{red}{
\begin{figure}[t]
    \centering
    \includegraphics[width=.5\textwidth]{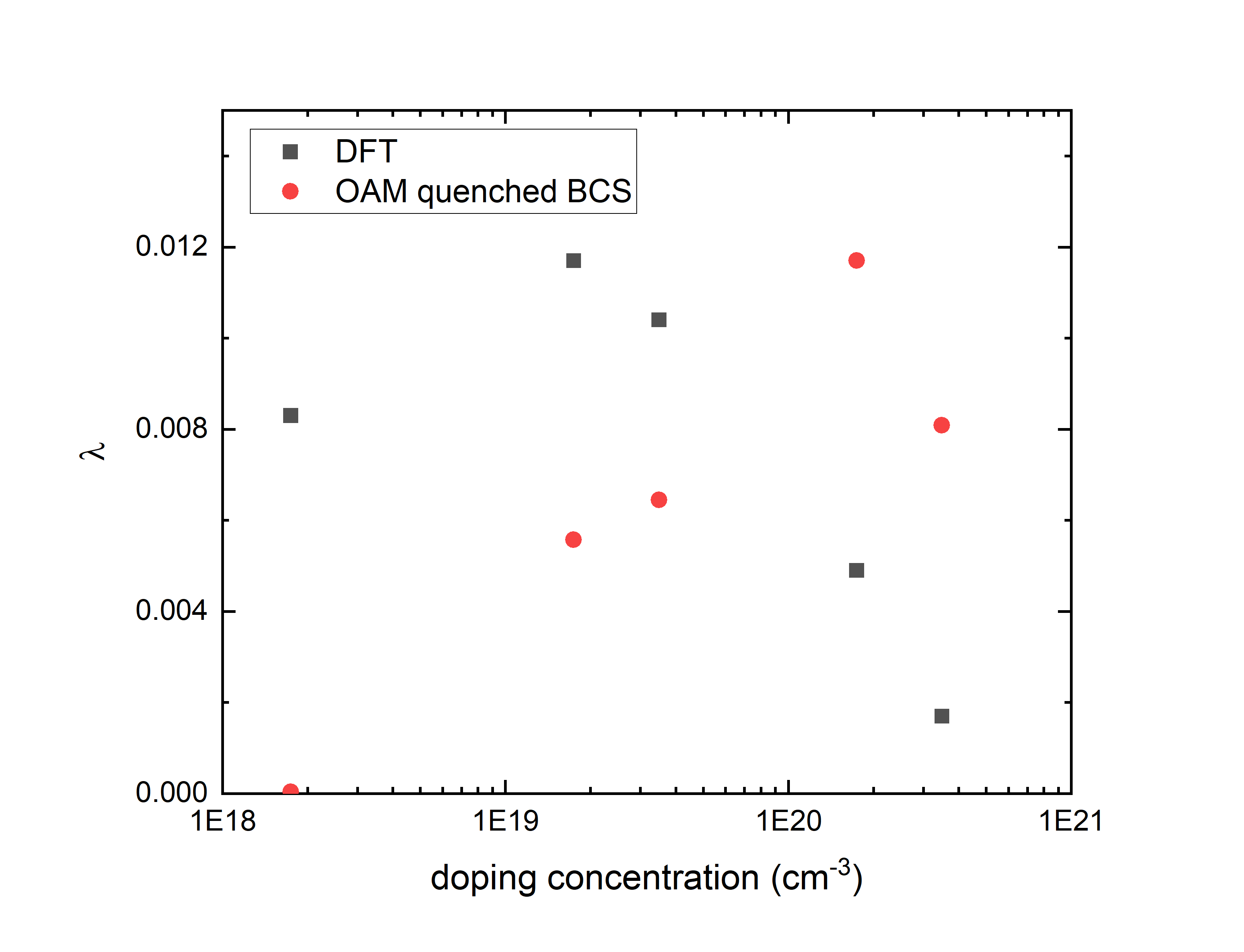}
    \caption{The BCS $\lambda$ that includes the orbital angular momentum quenching effect, marked red, is compared to the $\lambda$ obtained from our 
    first-principles calculation.}
    \label{FIG:OAMLambda}
\end{figure}
}

Given that our 
first-principles calculations fully incorporate the STO crystalline structure, Fig.~\ref{FIG:OAMLambda} plots a relevant comparison 
between our $\lambda$ and the BCS eigenvalue that incorporates the crystalline (and hence, orbital) symmetry effect. One such example is the recently discussed model~\cite{Gastiasoro2023} for the OAM quenching of the electron-phonon coupling 
at higher doping concentration. It has been pointed out previously \cite{Yu2022, Gastiasoro2022} that the electronic coupling to the polar phonons has strongest effect in the region close to the Ti $t_{2g}$ orbital energy crossing curves. But when the inter-orbital next-nearest neighbor hopping allowed by the orbital symmetry is accounted for, the resulting orbital hybridization quenches the OAM. This quenching effect as discussed in  Ref.~\cite{Gastiasoro2023} modifies the BCS pairing eigenvalues of Refs.~\cite{Yoon2021, Yu2022} 
into
\begin{equation}
\lambda_{\rm OAM} \propto \frac{n^{1/3}}{\omega^2}\frac{\sin^2(k_{F110}a)}{1+\left[\frac{4t_4 \sin^2(k_{F110}a)}{\xi}\right]^2},
\label{EQ:OAM}
\end{equation}
where 
the Fermi wave vector $k_{F110}$ in Fig.~\ref{FIG:OAMLambda} is determined by the same method as the $E_F$'s of Fig.~\ref{FIG:w2Evol}, $\omega$ the polar soft mode frequency at $\Gamma$ as plotted in Fig.~\ref{FIG:w2Evol}, $t_4$=40meV the inter-orbital next-nearest neighbor hopping between Ti $t_{2g}$ orbitals, and $\xi$=20meV is the Ti atomic spin-orbit coupling (details on the 
first-principles calculation of $t_4$ and $\xi$ is given in Appendix A); 
for convenience, the proportionality constant of Eq.~\eqref{EQ:OAM} is set so that the $\lambda_{\rm OAM}$  maximum would match the maximum of the DFT $\lambda$.
We see that both the optimal doping and the upper critical doping of the $\lambda_{\rm OAM}$ are 
larger, but within order of magnitude, from those of the $\lambda$ dome, making them 
closely match 
those of the $T_c$ dome shown in Fig.~\ref{FIG:Tc}. In the underdoped region, however, Fig.~\ref{FIG:OAMLambda} shows the difference between the $\lambda_{\rm OAM}$ dome and the $\lambda$ dome to be more substantial, 
in particular by the occurrence of the lower critical doping for the former around $n=1.7 \times 10^{18}$cm$^{-3}$.

\begin{figure}[t]
    \centering
    \includegraphics[width=.5\textwidth]{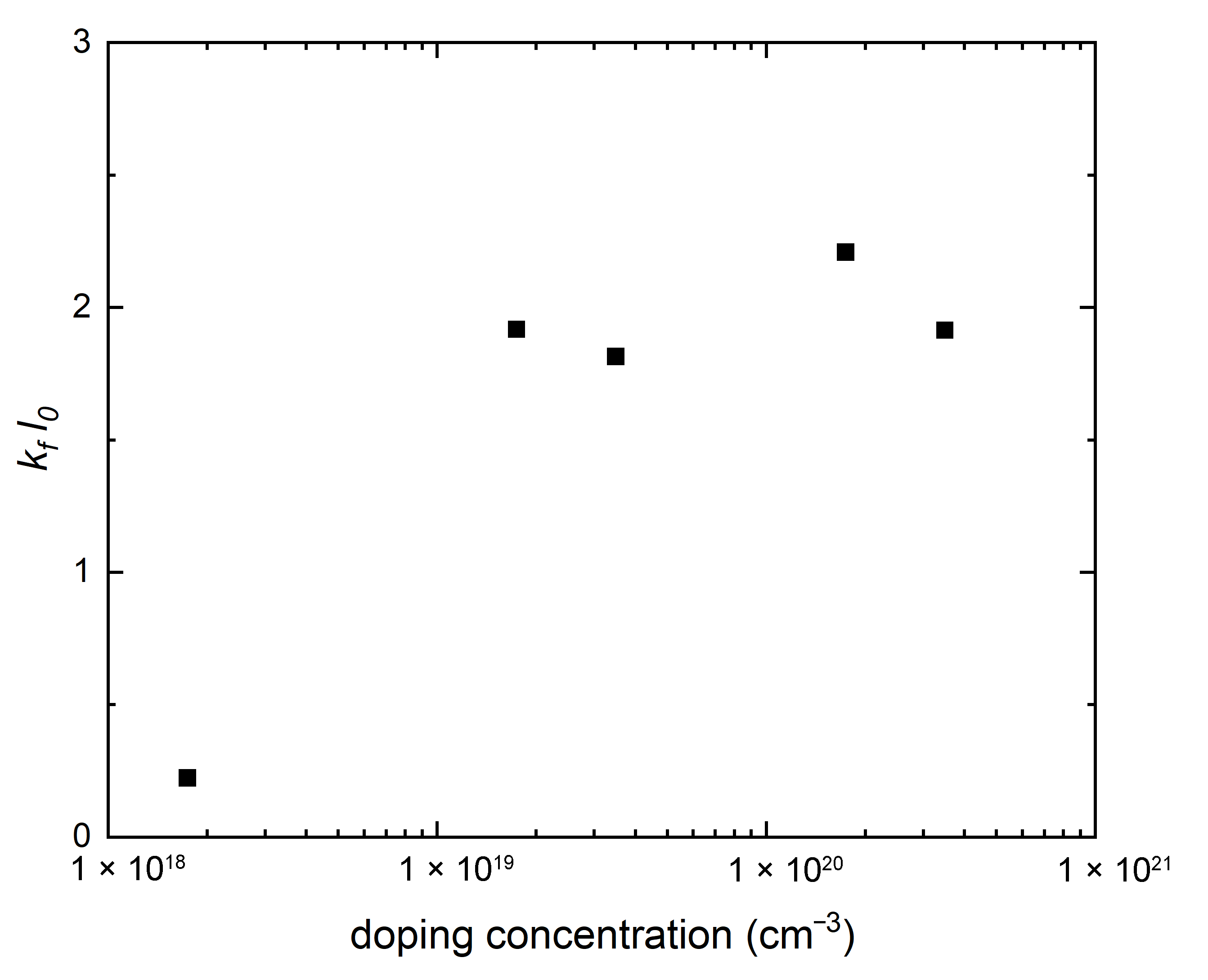} 
    \caption{The doping evolution of $k_F \ell_0$ where $\ell_0 \equiv v_p/\omega$ is the polar soft mode correlation length.}
    \label{FIG:ERange}
\end{figure}

Lastly, in Fig.~\ref{FIG:ERange}, we examine through the 
first-principles calculation the doping evolution of the product of $k_F$ and the polar soft mode correlation length $\ell_0$, which, as pointed out recently \cite{Fauque}, can tell us how many electrons are within the range of phonon-mediated interaction. 
Following Ref.~\cite{Fauque}, we concentrated on the interaction effect on the Fermi surface around the [110] direction where the linear electron-phonon coupling is expected to be strongest \cite{Gastiasoro2020, Yu2022, Gastiasoro2023}. This $\ell_0 \equiv v_p/\omega$ along the [110] direction is defined from the dispersion along the [110] direction $E^2(q)=v_p^2 q^2 + \omega^2 = \omega^2 (1+q^2\ell_0^2)$ of the mode with the strongest electronic coupling. As shown in Fig.~8 in Appendix B, our 
first-principles calculation shows the $\ell_0$ decreasing with doping, mainly due to $\omega$ increasing with doping, as shown in Fig.~\ref{FIG:w2Evol}, but also due to $v_p$ slightly decreasing with doping. As plotted in Fig.~\ref{FIG:ERange}, the doping evolution of $k_F \ell_0$,  
where $k_F$ is the $k_{F110}$ of Eq.~\eqref{EQ:OAM}, 
also shows a dome-like structure. Although the lower critical doping is clearly discernible for this dome, the optimal doping or the upper critical doping cannot be determined, as $k_F \ell_0$ essentially does not decrease in the overdoped region. Within the context of our calculation, the $\lambda_{\rm OAM}$ and the $k_F \ell_0$ domes of Figs.~\ref{FIG:OAMLambda} and \ref{FIG:ERange} clearly do not coincide, the former (latter) being shifted up (down) in the doping range  when compared to the experimental $T_c$ dome of Fig.~\ref{FIG:Tc}.

\section{Conclusions and Discussions}

We have shown in Figs.~\ref{FIG:MRLambda}  and \ref{FIG:EPhLambda} our 
first-principles calculation of the doping-dependent electron-phonon coupling. At all doping concentrations, the distribution of the mode-resolved electron-phonon coupling $\lambda_{{\bf q},\nu}$ as shown in Fig.~\ref{FIG:MRLambda} is strongly concentrated in the polar soft mode around $\Gamma$, a possible indication that these modes mediate the pairing interaction. Comparison between the dome we have found for the integrated electron-phonon coupling $\lambda$ shown in Fig.~\ref{FIG:EPhLambda} and the experimental $T_c$ dome shown in Fig.~\ref{FIG:Tc} shows a shift to a lower doping range of the former, with a significant enhancement in the underdoped region.

One possible upgrade on our results is to find a way to incorporate into our 
first-principles calculations the doping evolution of the energy splitting of the polar modes at $\Gamma$ between the longitudinal mode (LO1) and the transverse modes (TO1). According to infrared \cite{Vogt1981} and hyper-Raman \cite{Kamaras1995} experiments on the undoped STO bulk, the LO1 mode frequency at $\Gamma$ is greater than that of the TO1 mode by about 20.3meV, or equivalently 163cm$^{-1}$ according to the unit used in Fig.~\ref{FIG:MRLambda}. 
Given that at similar energy our Fig.~\ref{FIG:MRLambda} shows $\lambda_{{\bf q},\nu}$ to be insignificant, it may be reasonable to postulate that this splitting would suppress the electronic coupling to the LO1 phonons in the underdoped region. However, our DFT calculations 
ignored the LO1/TO1 energy splitting, and the resulting $\lambda_{{\bf q},\nu}$ is found to be approximately two orders of magnitude larger for the LO1 mode. This suggests that the large $\lambda$ in the underdoped region in Fig.~\ref{FIG:EPhLambda} would have been sharply attenuated upon inclusion of the LO1/TO1 splitting and that 
our calculation is more physically relevant in the overdoped region. 
It should be noted that experiments find the LO1/TO1 splitting to remain substantial ($\sim 10$meV) in the overdoped region, yet the upper critical doping of our Fig.~\ref{FIG:EPhLambda} approximates the upper critical doping of $T_c$ as shown in Fig.~\ref{FIG:Tc}; this may indicate the effect OAM quenching has on any polar displacements.

Lastly, it is worth noting that 
$k_F \ell_0$ is 
intrinsically independent of any orbital symmetry 
or any particular form of electron-phonon couplings, and consequently is potentially useful as diagnostic for non-linear electron-phonon coupling. Indeed, the two-phonon mediated pairing interaction \cite{Ngai1974, Kiselov2021, Volkov2022} would plausibly be more relevant for a larger $1/k_F \ell_0$, 
{\it i.e.} a larger number of phonons within the average distance between electrons; that would occur in the underdoped region according to our calculations for Fig.~\ref{FIG:ERange}. Devising a DFT calculation for the doping-dependent two-phonon coupling analogous to our linear electron-phonon coupling would be an interesting future challenge, and it would be worth comparing such calculations with the DFT $k_F \ell_0$ result.

\section*{Author Contributions}
M.P. performed numerical calculation. All authors contributed to designing the project and writing the manuscript.

\section*{Data Availability Statement}
Relevant data in this paper are available upon reasonable request.

\section*{Funding}
M.P and S.B.C. were supported by the National Research Foundation of Korea (NRF) grants funded by the Korea government (MSIT) (NRF-2023R1A2C1006144, NRF-2020R1A2C1007554, and NRF-2018R1A6A1A06024977).

\begin{acknowledgments}
We thank Hosub Jin and Sung-Hyon Rhim for useful discussions. We also thank the authors of Ref.~\cite{Yoon2021} for kindly granting permission to use a revised version of one of their figures as stated in the caption of Fig.~\ref{FIG:Tc}.
\end{acknowledgments}

\section*{Conflicts of Interest}
The authors declare no competing interests.

\begin{widetext}
\appendix

\section{STO electronic band structure fitting}
\label{scattering}

\begin{figure}[ht]
\centering
    \includegraphics[width=.67\textwidth]{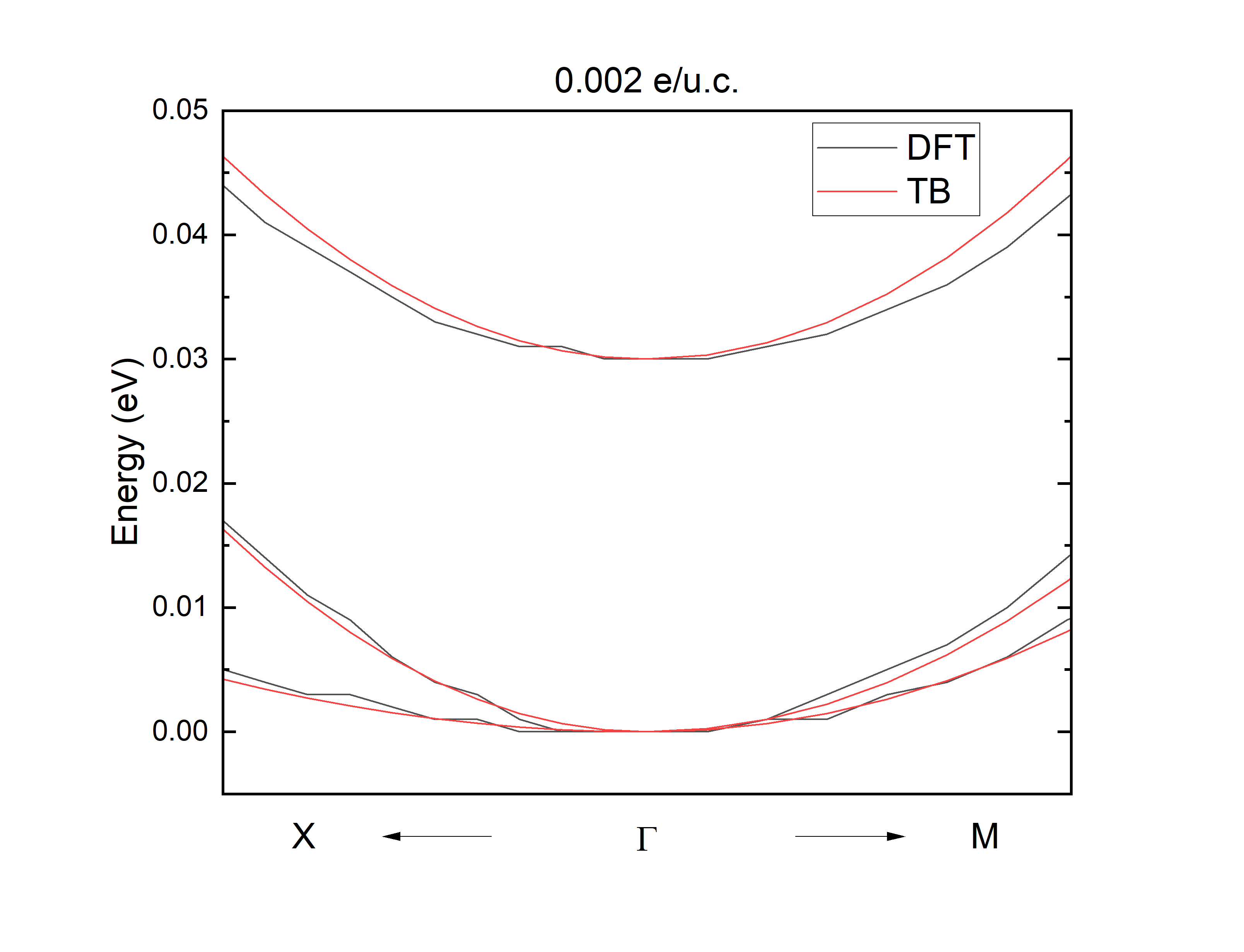} 
    \caption{Calculated band structure and tight-binding fitting near $\Gamma$ point with 0.002e/u.c.} 
    \label{FIG:tbfit}
\end{figure}

In a tight-binding model for the STO lowest conduction bands that includes hopping terms up to the next-nearest neighbor hopping, the eigenvalues of the matrix
\begin{eqnarray}
    \mathcal{H}_0({\bf k})&=&\left[\begin{array}{ccc}\epsilon_{xx}({\bf k}) & \epsilon_{xy}({\bf k}) & \epsilon_{xz}({\bf k})\\
    \epsilon_{yx}({\bf k}) & \epsilon_{yy}({\bf k}) & \epsilon_{yz}({\bf k})\\
    \epsilon_{zx}({\bf k}) & \epsilon_{yy}({\bf k}) & \epsilon_{zz}({\bf k})\end{array}\right]\nonumber\\
    &\,&+\frac{\xi}{2}\left[\begin{array}{ccc}0 & -i & 1\\ i & 0 & i\\ 1 & -i & 0\end{array}\right],
\end{eqnarray}
where
\begin{eqnarray}
\epsilon_{ij} &=& \delta_{ij}\left[2t_1\!\sum_{\mu\neq i}(1\!-\!\cos k_\mu) \!+\! 2t_2 (1\!-\!\cos k_i)\!-\!4t_3 \frac{\prod_\mu \cos k_\mu}{\cos k_i} \right]\nonumber\\
&\,& -4t_4 (1-\delta_{ij})\sin k_i \sin k_j
\end{eqnarray}
are the band energy eigenvalues of the Ti $t_2g$ basis $(|d_{yz}\,\sigma\rangle, |d_{zx}\,\sigma\rangle, |d_{xy}\,\bar{\sigma}\rangle)$, where $\sigma = \uparrow, \downarrow$ and $\bar{\sigma}$ signifies the spin state orthogonal to $\sigma$ \cite{Gastiasoro2020A, Gastiasoro2023}. 

We have found that the DFT electronic band dispersion is nearly independent of the doping concentration and, from the 0.002e/u.c. dispersion plotted in Fig.~S\ref{FIG:tbfit} above, can be fitted to the tight-binding parameters $t_i(i=1-4) = 515, 172, 74, 42$ meV and $\xi$ = 20 meV.

\section{The doping evolution of the polar soft mode velocity and correlation length}
\label{APP:phonon}

\begin{figure}[ht]
    \centering
    \includegraphics[width=.5\textwidth]{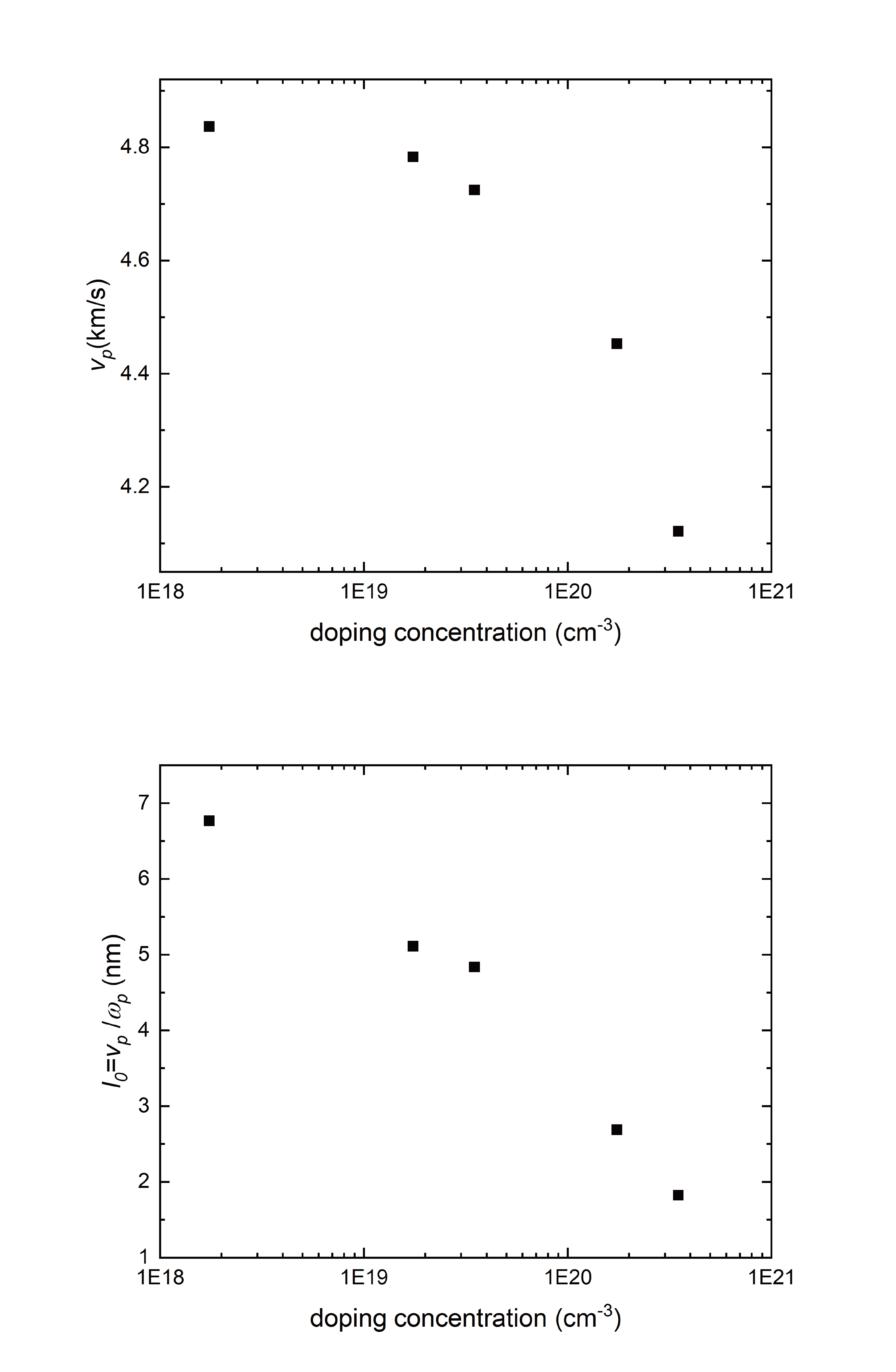}
    \caption{The doping evolution of the velocity $v_p$ (upper) and the correlation length $\ell_0 \equiv v_p/\omega$ (lower), respectively, of the polar soft mode. The former is obtained from fitting to $E^2 (q) = \omega^2 + v_p^2 q^2$ along the [110] direction, and the latter from $\omega$ taken from Fig.~2.}
    \label{FIG:phononV}
\end{figure}

\section{The doping evolution of the lattice constant}
\label{APP:phonon}

\begin{figure}[ht]
    \centering
    \includegraphics[width=.75\textwidth]{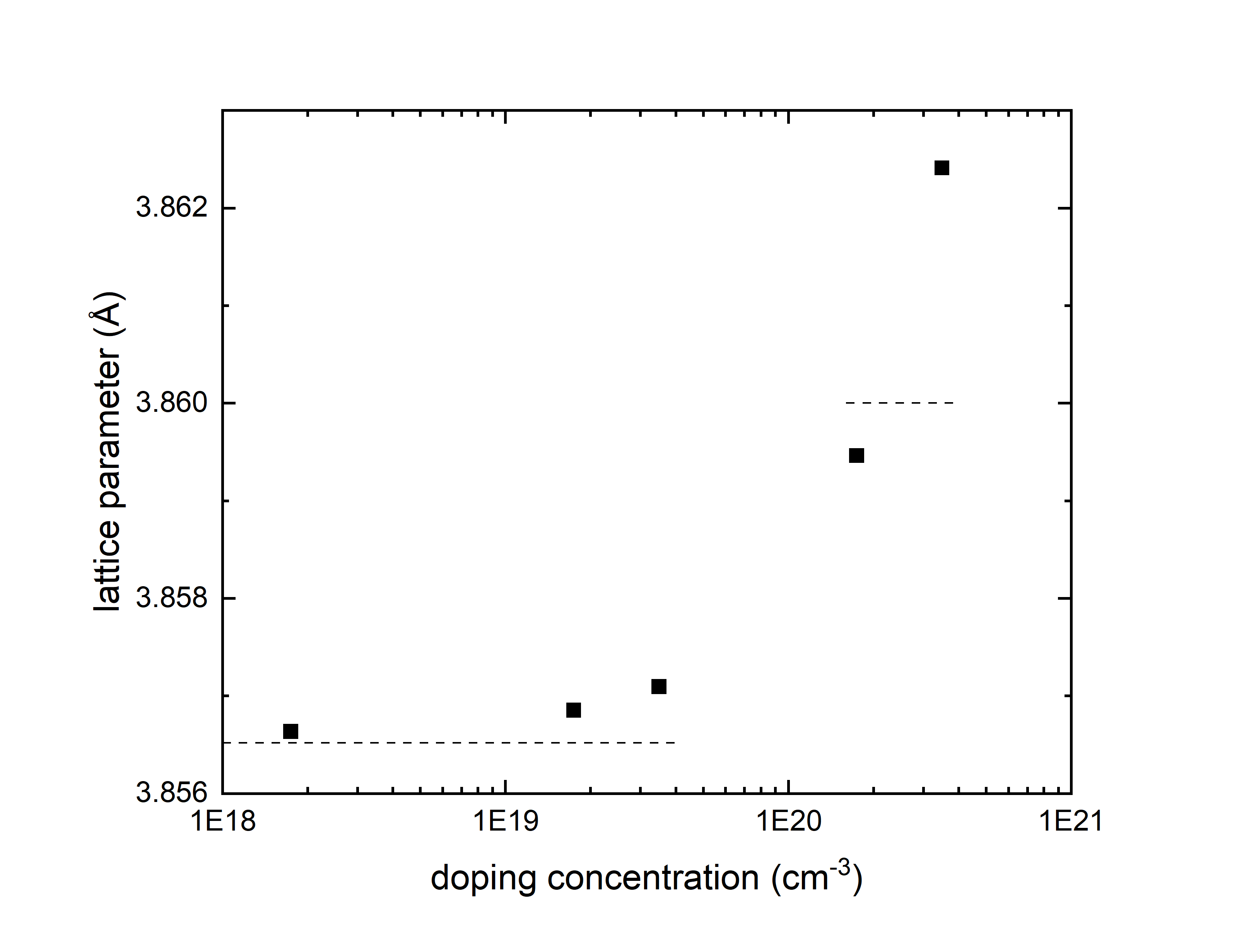}
    \caption{The doping evolution of the lattice parameter. As explained in Sec. II of the main text, we show the lattice constant for which the Hellman-Feynman force, {\it i.e.} stress, of the cubic lattice is minimized. For comparison, the values of $a_0=$3.8565 \AA~and $a_0=$3.8600 \AA~are indicated by the two sets of horizontal dashed segments.}
    \label{FIGlatticeV}
\end{figure}
\end{widetext}


\providecommand{\noopsort}[1]{}\providecommand{\singleletter}[1]{#1}%

\end{document}